\documentclass[%
 reprint,
 amsmath,amssymb,
 aip,
 rsi
]{revtex4-2}

\usepackage{graphicx}% Include figure files
\usepackage{dcolumn}% Align table columns on decimal point
\usepackage{bm}% bold math
\usepackage{tensor}
\usepackage{gensymb}
\usepackage{color}
\usepackage[dvipsnames]{xcolor}
\usepackage[version=4]{mhchem}
\usepackage{siunitx}
\DeclareUnicodeCharacter{FFFD}{\:{u}}
\DeclareUnicodeCharacter{0394}{$\Delta$}
\DeclareSIUnit\mbar{mbar}

\begin{document}
\newcommand{\dmsk}[1]{{\color{blue}{DMSK: #1}}}
\newcommand{\diego}[1]{{\color{green}{diego: #1}}}
\newcommand{\jack}[1]{{\color{red}{jack: #1}}}
\newcommand{\scott}[1]{{\color{Rhodamine}{scott: #1}}}
\bibliographystyle{unsrt}
\preprint{APS/123-QED}

\title{An Atomic Beam of Titanium for Ultracold Atom Experiments}% Force line breaks with \\

\author{Jackson Schrott}
    \affiliation{Department of Physics, University of California, Berkeley, CA 94720}
    \affiliation{Challenge Institute for Quantum Computation, University of California, Berkeley, CA 94720}
\author{Diego  Novoa}
    \affiliation{Department of Physics, University of California, Berkeley, CA 94720}
    \affiliation{Challenge Institute for Quantum Computation, University of California, Berkeley, CA 94720}
    \author{Scott Eustice}%
    \affiliation{Department of Physics, University of California, Berkeley, CA 94720}
    \affiliation{Challenge Institute for Quantum Computation, University of California, Berkeley, CA 94720}
\author{Dan M.\ Stamper-Kurn}
    \affiliation{Department of Physics, University of California, Berkeley, CA 94720}
    \affiliation{Challenge Institute for Quantum Computation, University of California, Berkeley, CA  94720}
    \affiliation{Materials Science Division, Lawrence Berkeley National Laboratory, Berkeley, CA 94720}

\date{\today}% It is always \today, today,

\begin{abstract}

We generate an atomic beam of titanium (Ti) using a ``Ti-ball'' Ti-sublimation pump, which is a common getter pump used in ultrahigh vacuum (UHV) systems. We show that the sublimated atomic beam can be optically pumped into the metastable $3d^{3}(^4F){4}s$ $a^5F_5$ state, which is the lower energy level in a nearly cycling optical transition that can be used for laser cooling.  We measure the atomic density and transverse and longitudinal velocity distributions of the beam through laser fluorescence spectroscopy. We find a metastable atomic flux density of \SI{4.3(2)e9}{\per\s\per\square\cm} with mean forward velocity \SI{773(8)}{\m/\s} at \SI{2.55}{\cm} directly downstream of the center of the Ti-ball. Owing to the details of optical pumping, the beam is highly collimated along the transverse axis parallel to the optical pumping beam and the flux density falls off as $1/r$. We discuss how this source can be used to load atoms into a magneto-optical trap.

\end{abstract}

\maketitle

%\tableofcontents

\section{Introduction}

Since the eras of Rabi and Ramsey, atomic beams have served as the starting point for a wide range of experiments investigating and harnessing atomic properties \cite{goldenberg_atomic_1961, rabi_atomic_1952, phil82}. For some elements, producing atomic beams in vacuum is straightforward.  For example, beams of alkali elements are easily produced by sublimation from samples placed in vacuum and heated to temperatures in the range of 300--\SI{600}{K}. However, many other elements require higher temperatures than what can be sustained by the walls of a vacuum chamber. Performing beam experiments with these elements requires in-vacuum heating of isolated samples with approaches such as laser ablation \cite{selter_analysis_1982}, high-temperature crucibles, or electron-beam bombardment  \cite{krzykowski_study_1997}.

In this work, we develop an atomic beam of Ti, a refractory metal relevant to precision spectroscopy and of emerging interest to quantum science. In terms of spectroscopic interest, Ti is a relatively light transition metal in which the atomic structure of open $d$-shell atoms can be studied at moderate complexity, allowing measurements to be compared with tractable theoretical calculations \cite{safronova_all-order_2008, neel21isotope, furm96ti}.  Previous experiments have also focused on spectroscopic probes of the $^{50}$Ti nucleus, which is of interest to nuclear physics due to its combination of being near the magic proton number, $Z\!=\!20$ and having the magic neutron number, $N\!=\!28\,\,$ \cite{jin_hyperfine_2009,lombard_size_1990, angeli_table_2013}.

\begin{figure}[t]
    \centering
    \includegraphics[width=\linewidth]{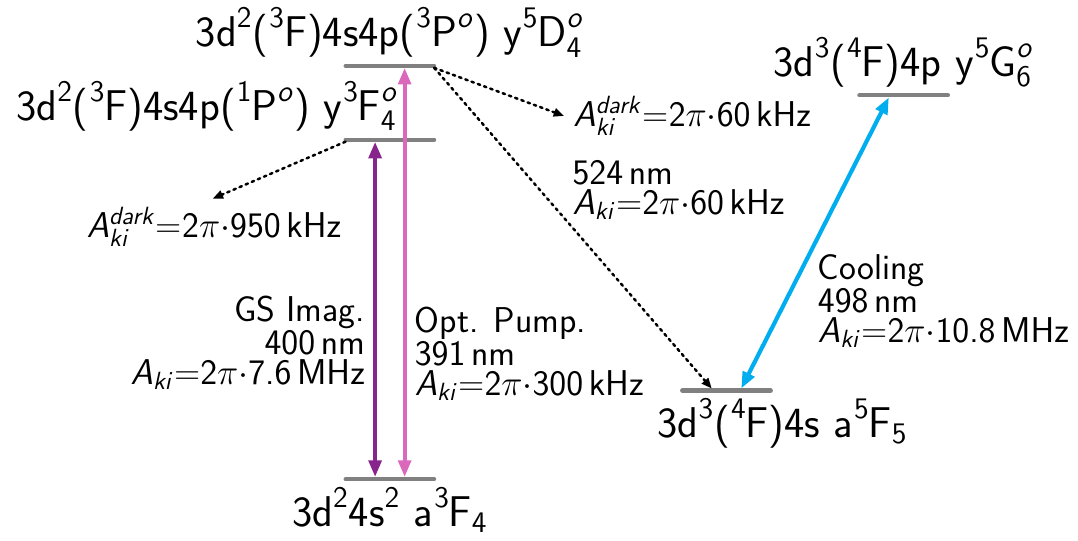}
    \caption{Energy levels and transitions in atomic Ti relevant to this work. Light at \SI{391}{\nm}, \SI{400}{\nm}, and \SI{498}{\nm} wavelengths are used for optical pumping, ground state imaging and metastable state imaging respectively. The \SI{524}{\nm} line shows the wavelength of the emitted photon for atoms that are optically pumped to the $a^5F_5$ state.}
    \label{fig:level_diagram}
\end{figure}

Our chief motivation for developing an atomic beam of Ti is to produce an atom source for laser cooling and trapping experiments \cite{eustice_laser_2020}. As shown in Fig.\ \ref{fig:level_diagram}, laser cooling can be realized on a broad-linewidth transition at 498 nm wavelength from the $3 d^3 (^4F) 4s$ $a^5F_5$ state (herein referred to as the metastable or laser-cooling state) to the $3 d^3 (^4F) 4p$ $y ^5G_6^o$ state. This $J \rightarrow J+1$ transition is nearly cycling, with a calculated dark state branching ratio of $\sim3 \times 10^{-7}$ (see supplemental information of Ref.\ \cite{eustice_optical_2023}), satisfying requirements for both laser cooling and high-fidelity optical detection.  Laser-cooled Ti may find applications as both a telecommunications-wavelength optical clock\cite{eustice_optical_2023} and a platform for quantum simulation of interesting many-body physics.

In producing an atomic beam of Ti for use in laser cooling experiments, we face two challenges. First, Ti's vapor pressure remains below \SI{2e-8}{\mbar} at temperatures up to \SI{1350}{\kelvin}, so high temperatures are required to produce a bright source of atoms \cite{edwards1953vapor}.  Next, the high energy of the $a^5F_5$ state --- $E \!=\! h c \times \SI{6843}{\per\centi\m} \!=\! k_\mathrm{B} \times \SI{9846}{\kelvin}$  --- dictates that a Ti beam produced by sublimation at \SI{1350}{\kelvin} will have only a small population ($\sim\! 0.04\%$) in the laser-coolable state. Therefore, the second challenge is to enrich the atomic population in the metastable state.

We have produced a bright source of Ti atoms in their metastable state using a commercially available titanium sublimation vacuum pump (Ti-sub). The Ti-sub provides a high flux source of ground state atoms, and optical pumping (OP) enriches the metastable state population. Operating the Ti-sub at a measured temperature of \SI{1345(30)}{\kelvin}, we find a metastable atomic flux density of \SI{4.3(2)e9}{\per\s\per\square\centi\m} at a distance of \SI{2.55}{\centi\m} downstream of the center of the Ti-sub. The resulting beam has a divergence half angle of $\sim\!0.6\degree$ along the direction of the OP beam and $\sim\!5\degree$ in the other transverse direction. As discussed in Section \ref{sec:characterization}, the small divergence along the OP direction is a consequence of the narrow linewidth of the OP transition and the latter divergence is purely due to the geometry of the OP beams in our set up.

In a typical Ti-sub pump, a titanium-molybdenum filament is mounted in vacuum and heated resistively, causing the Ti to sublimate and coat the walls of the vacuum chamber. The exposed coating of Ti acts as a chemical getter, absorbing residual gases, and pumping the chamber to a lower pressure.  An alternative version of the Ti-sub filament is the so-called Ti-ball, where current is run through a tungsten filament that is nestled beneath a Ti shell. In all of the following, a Ti-ball style Ti-sub pump was used. Here we are interested in the sublimated Ti atoms as an atomic beam. In related work, the emission from a Ti-sub pump was adapted as a source for molecular-beam epitaxy \cite{thei96ti}.

Our paper is structured as follows.  In Section \ref{sec:tiball} we discuss the hardware construction of our beam source.  Section \ref{sec:detection} describes our methods for detecting the emission of the beam source, both for atoms in the ground-state term ($a ^3F_4$) and also in the metastable laser-cooling  state ($a ^5F_5$).  In Section \ref{sec:op} we summarize our scheme for optically pumping atoms into the metastable state.  Section \ref{sec:characterization} presents our data that characterizes the atomic beam source.  We demonstrate that the Ti-ball produces a strong beam of ground-state atoms that is well characterized by a Maxwell-Boltzmann velocity distribution.  We observe a weak metastable beam produced directly from the Ti-ball, and then find that the metastable population can be enhanced greatly through optical pumping.  This optical pumping selects a narrow class of transverse velocities, but produces an atomic beam with a broad longitudinal velocity distribution.  We conclude in Section \ref{sec:conclusion} with some discussion of using this beam source to load magneto-optical traps. A detailed discussion of how fluorescence counts were used to extrapolate atom density is left to Appendix \ref{sec:appendix}.%\jack{added this line}

\section{Atomic source}
\label{sec:tiball}

\begin{figure}[h]
    \centering
    \includegraphics[width = 3 in]{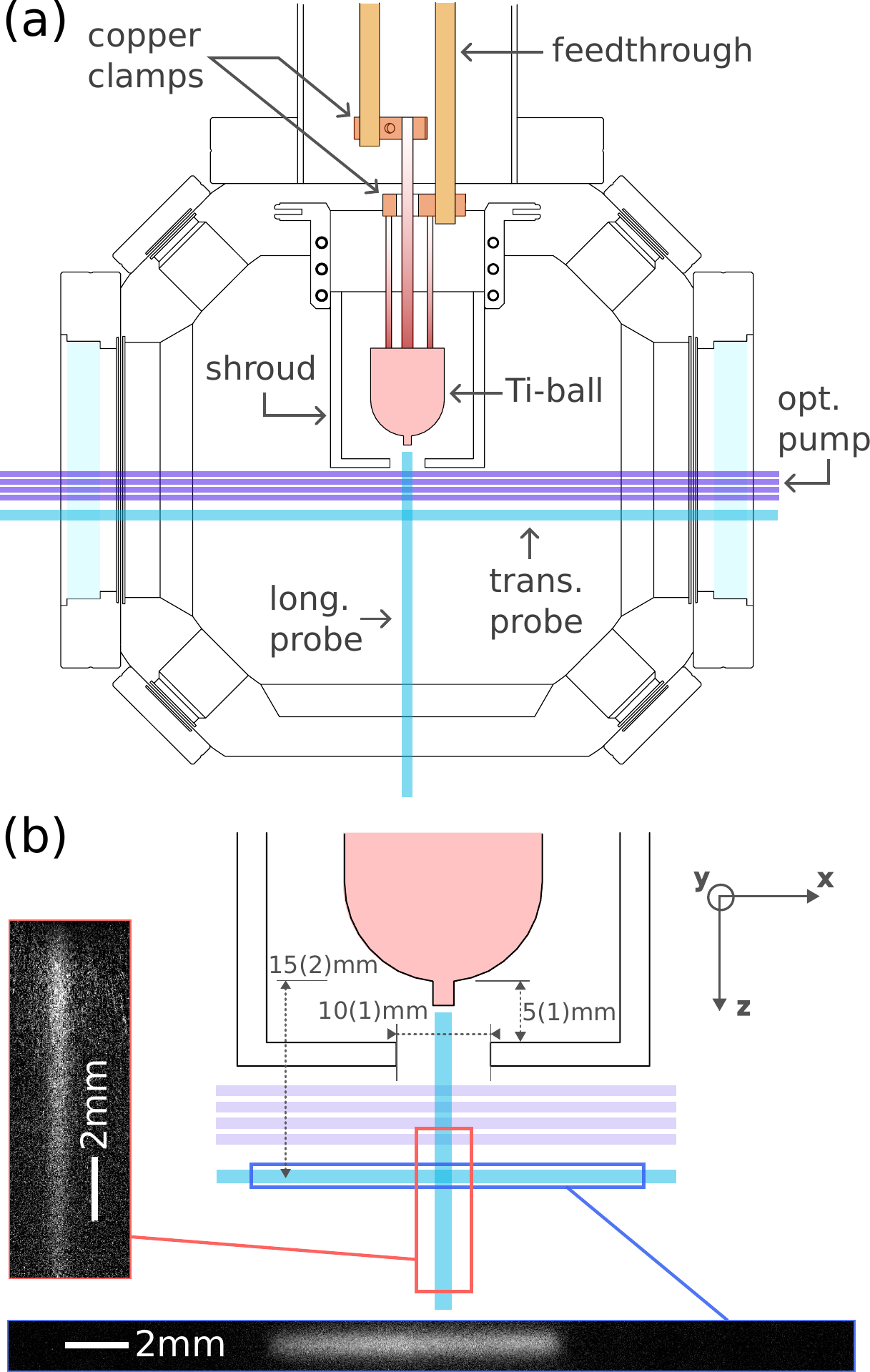}
    \caption{(a) Schematic of the experimental apparatus. The Ti-ball is attached via copper clamp to a copper UHV electrical feedthrough. Ground- or metastable-state imaging light (teal) is sent in along the $x$ or $-z$ axis. OP light (purple) is sent in along the $x$-axis. A CF 4.5 inch tee housing the the longitudinal viewport, an ion pump, and a UHV valve is hidden for clarity. (b) A magnified view of the area where the Ti beam interacts with the lasers. Representative images of the transverse and longitudinal beam are shown, with a scale bar indicated at the bottom left. The imaging system (lenses and camera) lies below the chamber (into the page). }
    \label{fig:apparatus}
\end{figure}
Fig.\ \ref{fig:apparatus} shows the experimental apparatus. An Agilent mini Ti-ball (Part Number 9160008) \cite{noauthor_mini_nodate} is attached to a conflat (CF) UHV electrical feedthrough with custom built copper clamps. The feedthrough conductor pins are copper and have in-vacuum lengths of \SI{8}{\cm} and \SI{10}{\cm}. The pins are \SI{0.64}{\cm} in diameter and the total thermal resistance of the feedthrough is approximately $R_{\theta}\!=\!\SI{3.5}{\kelvin/\watt}$.

The Ti-ball is heated using dc current up to \SI{42}{\ampere}, which corresponds to \SI{270}{\watt} of electrical power dissipated by the tungsten heater within the Ti-ball. Based on known curves for the vapor pressure of Ti vs.\ temperature, and the evaporation rate from the Ti-ball vs.\ current specified by the manufacturer \cite{noauthor_mini_nodate, yaws_yaws_2015}, the expected temperature of the Ti-ball at this power is $\sim\!\SI{1600}{\kelvin}$. However, as discussed in Section\ref{sec:characterization}, we measure a temperature of \SI{1345(30)}{\kelvin}.  We attributed the lower observed temperature to the fact that our feedthrough has a higher thermal conductance than the one used by the manufacturer. Documentation provided by the manufacturer suggests the Ti-ball can be operated at \SI{270}{\watt} for over 2000 hours before it is depleted, however the sublimation rate associated with the lower temperature we observe suggests a lifetime which is at least two orders of magnitude longer than this.

The tip of the Ti-ball protrudes \SI{60(2)}{\mm} into a UHV chamber (Kimball Physics 6" spherical-square, part number 53-140320), measured from the CF flange of the chamber, and is surrounded by a hollow \SI{19}{\mm} radius Ti cylinder (``shroud'') which has an aperture of \SI{5}{\mm} radius on its front face. The Ti-ball is recessed \SI{5(1)}{\mm} behind the aperture of the shroud, measured from the most prominent point of the Ti-ball hemisphere.  The purpose of the shroud is to eliminate direct line of sight from the Ti-ball to vacuum viewports to avoid coating by an opaque layer of Ti.  The aperture provides little collimation, with the emitted Ti beam diverging with an opening angle of around 30\degree.

\section{Fluorescence detection of the atom beam}
\label{sec:detection}

Atoms in the beam may be imaged and their velocities measured by way of fluorescence spectroscopy. A probe laser interacts resonantly with unpolarized atoms in the beam, causing them to scatter photons isotropically. The scattered light is imaged onto a camera, and its intensity used to measure the local density of atoms. 

\begin{figure}[t]
    \centering
    \includegraphics[width=\linewidth]{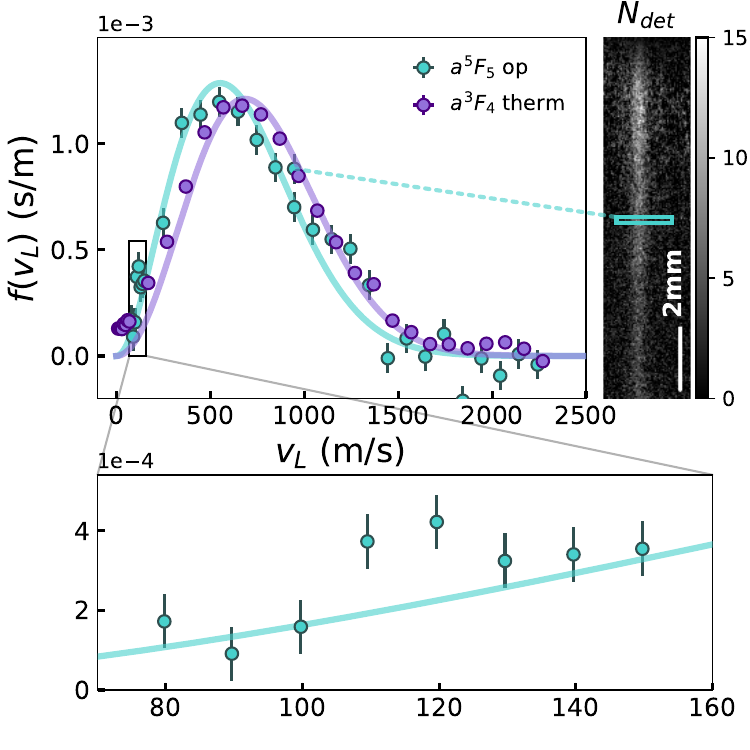}
    \caption{Longitudinal Doppler profiles for ground-state (purple) and optically pumped metastable state (teal) atoms.  The profiles are normalized to 1. The solid purple line shows a Maxwell-Boltzmann fit to the $a^4F_4$ data, giving a fit temperature of $T\!=\!\SI{1345(30)}{\kelvin}$. The solid teal line shows the same Maxwell-Boltzmann distribution multiplied by a function that models the efficiency of optical pumping as a function of forward velocity.  These data (highlighted in the inset figure) confirm the presence of a low-velocity population of metastable-state atoms, suited for loading a magneto-optical trap either directly or with the aid of a Zeeman slower. Shown to the right of the Doppler profiles is a representative image and the region of interest in which we sum fluorescence counts.}
    \label{fig:LDPs}
\end{figure}

Two orientations of probe beams were used in this study: one propagating transverse to the emitted beam, and one propagating toward and longitudinal to the emitted beam.  Fig.\ \ref{fig:apparatus}b shows typical images of scattered light from both transverse (along the $x$-axis in Fig.\ \ref{fig:apparatus}) and longitudinal (along the $-z$-axis in Fig.\ \ref{fig:apparatus}) probes. By quantifying the fluorescence in either situation, we characterize the beam's velocity distributions, $f(v_T)$ and $f(v_L)$, along the transverse and longitudinal directions, respectively. A detailed derivation of how fluorescence counts are converted to atom density is provided in Appendix \ref{sec:appendix}.

We detect atoms in two different electronic states. First is the $3d^24s^2$ $a^3F_4$ fine-structure level (referred to hereafter as the ground state), which is located \SI{387}{\cm^{-1}} above the true lowest energy fine structure level, and is highly populated in the sublimated gas. The second is the aforementioned metastable  $3d^34s^2$ $a^5F_5$ state which is scarcely populated at \SI{1345}{\kelvin}. As shown in Fig.\ \ref{fig:level_diagram}, atoms in these two states can be made to fluoresce by probe lasers at \SI{400}{\nm} and \SI{498}{\nm} wavelength, respectively. The \SI{400}{\nm} wavelength light drives the transition between the $a^3F_4\rightarrow y^3F_4$ states, which has an Einstein $A$ coefficient of $A_{ki}\!=\!2\pi\times\SI{7.6}{\MHz}$ and a branching ratio to dark states of $\sim\!0.1$.  Owing to this factor, and the fact that it is a $J\rightarrow J$ transition with a dark $m_J$ state, the average $a ^3F_4$ atom may scatter about 5 photons at \SI{400}{\nm} wavelength before being pumped to a dark state and no longer fluorescing. The \SI{498}{\nm} wavelength light drives the laser cooling transition between the $a^5F_5\rightarrow y^5G_6$ states, which has an Einstein coefficient of $A_{ki}\!=\!2\pi\times\SI{10.8}{\MHz}$ and a branching ratio to dark states of $\sim\! 10^{-7}$.  Given this small branching ratio, a metastable state atom will continue to fluoresce \SI{498}{\nm} wavelength light as long it remains within the fluorescence laser beam, or until the mechanical action of the light pushes the atom out of resonance\cite{eustice_optical_2023}.  The \SI{400}{\nm} wavelength  light is generated by an external-cavity diode laser  (TOptica DLC DL Pro) and the \SI{498}{\nm} wavelength light is generated by frequency-doubling the output of a titanium-sapphire laser (MSquared SolsTiS and ECD-X).

The fluorescence of the Ti atoms is imaged through a 0.17(2) numerical aperture (NA) imaging system below the chamber (into the page on Fig.\ \ref{fig:apparatus}a). The magnification of the imaging system is 0.17(1) and the total light collection efficiency of the system including NA, transmission through optics, and quantum efficiency is $\epsilon_{498}\!=\!2.6(2) \times 10^{-3}$ at \SI{498}{\nm} wavelength, and $\epsilon_{400}\!=\!1.8(2) \times 10^{-3}$ at \SI{400}{\nm} wavelength.  A Thorlabs Zelux camera (model no. CS165MU1) was used.

\begin{figure}[!htb]
    \centering
    \includegraphics[width=\linewidth]{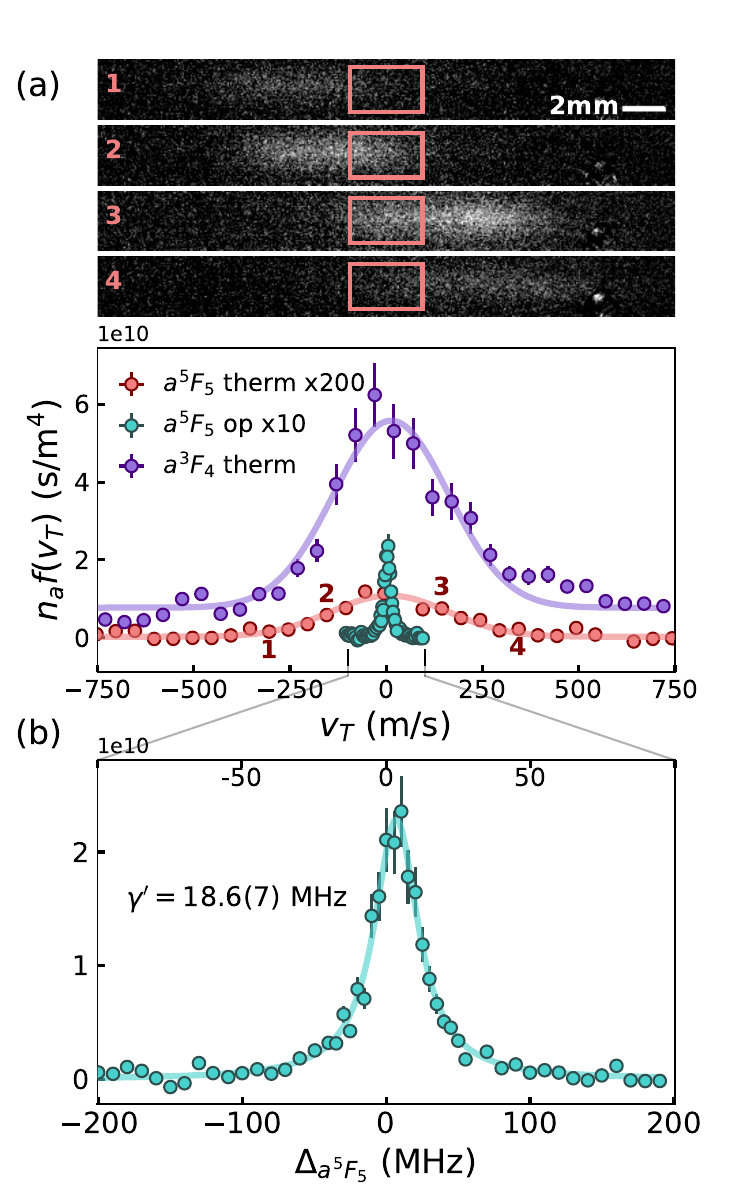}
    \caption{(a) Transverse Doppler profiles of thermal (unpumped) \ce{a^3F_4} (purple), thermal \ce{a^5F_5} (red), and optically pumped \ce{a^5F_5} (teal) atoms. The thermal and optically pumped metastable profiles are multiplied by factors of $\times200$ and $\times10$ for visibility. Shown above are characteristic fluorescence images. Pixels in the highlighted region of interest are summed and converted into atom density per velocity bin as plotted below. The propagation direction of the atomic beam is from the top of the images to the bottom. The four images show that as the detuning of the probe laser is scanned, the probe interacts with atoms in different regions of the beam. (b) Enhanced view of the optically pumped transverse Doppler profile. The half-width half maximum peak width of $\gamma^\prime\! =\!  \SI{18.7(7)}{\MHz}$ corresponds to a transverse velocity spread of $\Delta v_{t, op}\!=\!\SI{7.1}{\m/\s}$, showing artificial collimation of the beam. Owing to uncertainties in transition rates, the metastable-state data carry an additional $4\%$ systematic uncertainty and the ground-state data carry an additional $7\%$ uncertainty}
    \label{fig:TDPs}
\end{figure}

\section{Optical pumping scheme}
\label{sec:op}

Our setup also makes use of light at a wavelength of \SI{391}{\nm} to drive a transition from the $a ^3F_4$ ground state to the $3 d^2(^3F) 4s 4p (^3P^o) y ^5D_4^o$ excited state (see Fig.\ \ref{fig:level_diagram}). Excitation on this transition serves to pump the atoms optically.  We discern three different decay outcomes from the excited state: an atom may return to the ground state (Einstein $A$ coefficient $A_\mathrm{ground}\!=\!2 \pi \times \SI{287(127)}{\kHz}$), decay to the metastable state ($A_\mathrm{pump} \!=\! 2 \pi \times \SI{63(63)}{\kHz}$), or decay to a different metastable state where it is no longer coupled to OP light ($A_\mathrm{dark} \!=\! 2 \pi \times \SI{63(17)}{\kHz}$)\cite{eustice_optical_2023}.  Given these decay rates, a ground-state atom will scatter on the order of $A_\mathrm{ground} / (A_\mathrm{pump} + A_\mathrm{dark}) \!=\! 3$ photons before being optically pumped, with the pumping efficiency after many photon scatterings to the $a ^5F_5$ state being $P_\mathrm{max} \!=\! A_\mathrm{pump} / (A_\mathrm{pump} + A_\mathrm{dark}) \!=\! 0.5$.

The OP transition is rather narrow, with a natural linewidth of $\gamma_\mathrm{op} \!=\! 2 \pi \times \SI{414(140)}{\kHz}$ \cite{eustice_optical_2023}.  The transition from the ground state to the optical-pumping excited state, linking a spin-triplet to spin-quintuplet state, is weakly allowed owing to state-mixing in the excited state manifold.  The decay on the spin-allowed transition to the spin-quintuplet metastable state is weak because of the small matrix element for transitions from the $4 p$ to the $3 d$ orbital.

Light at the wavelength of \SI{391}{\nm} is generated from a Littrow-configuration external cavity diode laser (MOGLabs model LDL).  As shown in Fig.\ \ref{fig:apparatus}, this light transversely illuminates the atomic beam.  The light can be passed either once or multiple times through the atomic beam to increase the OP efficiency.  Optical access limitations of our setup permit the beam to be reflected up to four times through the atomic beam; straightforward improvements to optical access should allow for several more reflections.  At closest approach, the OP light passes 2 mm from the face of the Ti-ball shroud.

The beam has waists ($1/e^2$-intensity radius) of \SI{448.9(1)}{\um} along the $y$-axis in Fig.\ \ref{fig:apparatus} and \SI{178.2(1)}{\um} along the $z$-axis (longitudinal atomic beam direction).  With as much as \SI{20}{m\watt} of power available on the OP transition, we can thus provide peak intensities as high as \SI{16}{\watt/\square\cm}. This peak power is three orders of magnitude larger than the saturation intensity $I_\mathrm{sat,op} \simeq 4\, \mbox{mW}/\mbox{cm}^2$ for this transition.  Owing to this high power, we expect the OP transition to be power-broadened significantly.  As discussed below, this effect broadens the range of transverse velocities at which we can optically pump our beam.

Taking $\bar{v} \!=\! \sqrt{8 k_B T / \pi m} \!=\!\SI{773}{\m/\s}$ as the mean longitudinal thermal beam velocity, with $T \!=\!\SI{1345}{\kelvin}$ being the Ti-ball temperature and $m \!=\! \SI{48}{amu}$ the mass of the most abundant $^{48}$Ti isotope, an atom at $\bar{v}$ passes through the \SI{0.18}{\mm} beam radius in about \SI{0.2}{\us}, somewhat shorter than the natural lifetime (\SI{0.4}{\us}) of the $y ^5D_4$ state.  Thus, even at very high power, such a fast atom may scatter only one photon from one pass of the OP light.  While we can ensure that several photons may be scattered by using multiple passes of OP light, we suspect our setup is still sub-optimal for optically pumping the fast atoms within the atomic beam.  However, considering the specific application of our setup toward laser cooling of atomic Ti, we are most interested in optically pumping atoms at longitudinal velocities lower than $\bar{v}$, perhaps even only those below the $\sim30-\SI{100}{\m/\s}$ capture velocity of a magneto-optical trap.  For these slower atoms, the finite transit time through the OP beams does not to degrade the efficiency of optical pumping.

\section{Beam characterization}
\label{sec:characterization}
Fig.\ ~\ref{fig:LDPs} shows results of longitudinal fluorescence probes of the atom beam. In each of the datasets shown, we step the laser detunings, $\Delta$, over a range corresponding to the plotted Doppler velocities.  At each detuning, we average 5 fluorescence images like those found at the right of the figure. The number of photons detected in the highlighted region of interest during the exposure time is used to extract $f(v_L)$ per an equation analogous to Eq.\ \ref{eqn:Nfparallel}.  We are able to measure $f(v_L)$ both for the $a^3F_4$ ground-state beam and for the optically pumped $a ^5F_5$ metastable-state beam.  The signal from the unpumped, thermally produced metastable-state beam was too weak to observe.

Because we probe only a small ray of the atom beam, the thermal beam of $a^3F_4$ atoms is expected to have a Maxwell-Boltzmann longitudinal velocity distribution, of the form
\begin{equation}
    f(v_L) = \frac{32 v_L^2}{\pi^2\bar{v}^3} \exp(-4v_L^2/\pi \bar{v}^2).
\end{equation}
where $\bar{v}\!=\!\sqrt{8k_BT/\pi m}$.  Fitting this function to our $a^3F_4$ beam data, we determine the mean forward velocity to be $\bar{v}\!=\!\SI{773(8)}{\m/\s}$, corresponding to a temperature of $T\!=\!\SI{1345(30)}{\kelvin}$.

In the optically pumped beam, the population of \emph{fast atoms} is suppressed because their finite transit time in the pumping beam reduces the OP efficiency.  We see this effect qualitatively by the observation that the peak signal from optically pumped metastable atoms is found at lower longitudinal velocities than that from thermal ground-state atoms. The distribution of these atoms can be predicted by multiplying the distribution for ground state atoms by a function which takes into account the efficiency of optical pumping as a function of interaction time.  As shown in Fig.\ \ref{fig:LDPs}, this pumping-efficiency-modified distribution fits our data well, supporting our characterization of the OP process.

We highlight the fact that the optically pumped beam contains an ample population of atoms at low velocities, e.g.\ those below the $\sim 100$ m/s typical capture velocity of a magneto-optical trap (see inset of Fig.\ \ref{fig:LDPs}).  Thus, this optically pumped beam is suitable for directly loading magneto-optical traps.

Next, we examine the transverse velocity distribution of the atomic beam, shown in Fig.\ \ref{fig:TDPs}. The poor collimation of the atomic beam in our setup is evidenced by the large longitudinal velocity distribution of the beam, even when detected within a small region of interest within the transverse fluorescence image.  From Gaussian fits to the frequency variation of the fluorescence data, we extract an rms transverse velocity width of $\sigma^{a^3F_4}_T\!=\!\SI{150(10)}{\m/\s}$ for the ground-state and of $\sigma^{a^5F_5}_{T,\mathrm{therm}}\!=\!\SI{152(11)}{\m/\s}$ for the unpumped, thermally produced metastable atomic beams.

The signal from the thermal metastable-state beam is far weaker than that of the ground-state beam (note the x200 factor in comparing data for these two populations in Fig.\ \ref{fig:TDPs}a).  The ratio of atomic density in the $a ^5F_5$ metastable state and $a ^3F_4$ ground state is, presumably, determined by the Boltzmann factor that accounts for the energy difference between these two states, the temperature of the sublimating Ti-ball source, and the degeneracy of the two states. Using the longitudinal fit temperature of \SI{1345}{\kelvin}, we find an expected population ratio of $\frac{11}{9} \exp( - \Delta E/k_B T)=1.27\times10^{-3}$. From integrating the Gaussian fits in Fig.\ \ref{fig:TDPs}a, we ascertain from our fluorescence data a ratio of $\sim1.1(2)\times10^{-3}$, in excellent agreement with expectation.

From Boltzmann statistics, the fractional population in the $a^3F_4$ state at \SI{1345}{\kelvin}. Considering that from the position of the detected fluorescence volume, the Ti-ball looks like a uniform Lambertian emitter, we surmise a total sublimation rate of $n_a\bar{v}\, m\!=\! \SI{1.8(2)e-5}{\g\per\s\per\square\m}$, which is about a factor of 10 higher than expectation for a Ti surface at \SI{1345}{\kelvin}, and is consistent with a temperature of \SI{1430}{\kelvin}.  This discrepancy may suggest incorrect accounting of the collection efficiency of our imaging system or other systematic effects which shift the overall normalization of the data, but not the fitted temperature or ratio of the integrals of the curves.  Given the surface area of the Ti-ball,  the total sublimation rate we deduce is \SI{0.10(2)}{\mg/\hour}, considerably lower than the \SI{6}{\mg/\hour} specified by the manufacturer at \SI{42}{\ampere} running through the Ti-ball. As mentioned, we expect the lower observed temperature to be the consequence of our copper feedthrough having a larger thermal conductance than the steel feedthrough designed by Agilent.
 
Let us now compare the flux of metastable state atoms produced with (teal data in Fig.\ \ref{fig:TDPs}) and without (red data in Fig.\ \ref{fig:TDPs}) OP light. The transverse probe of the optically pumped metastable beam reveals two striking features.  First, we find that optical pumping enhances the population of metastable state atoms, but only within a narrow range of transverse velocities.  As shown in Fig.\ \ref{fig:TDPs}, while the fluorescence of the unpumped metastable beam is a broad function of probe-light detuning $\Delta$, the fluorescence of the pumped beam varies sharply with probe detuning.  We apply a Lorentzian fit to the pumped beam data, finding a half width at half maximum of $\gamma^\prime \!=\! 2 \pi \times \SI{18.7(7)}{\MHz}$.  Empirically relating ${\gamma^\prime}^2 \!=\! \gamma^2(s+1) + k^2 (\Delta v_{t,op})^2$, with $\gamma$ being the natural linewidth of the laser-cooling transition and $k$ the wavevector of 498-nm-wavelength light, we find a transverse velocity width of $\Delta v_{t,op} \!=\! \SI{7.1}{\m/\s}$ for the optically pumped beam.  

This narrow velocity selection reflects the narrowness of the power-broadened linewidth of the \SI{391}{\nm} OP transition as compared to the large transition Doppler width of the poorly collimated Ti-ball emission. The measured width $\gamma^\prime$ agrees with our order-of-magnitude expectation for power broadening in the OP beam. 

\begin{figure}
    \centering
    \includegraphics[width=\linewidth]{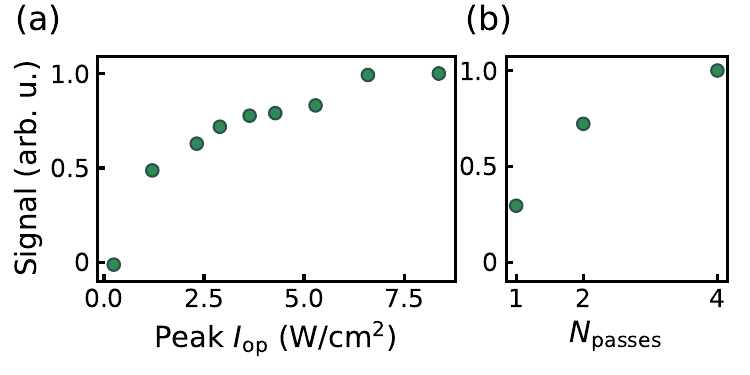}
    \caption{Variation of the metastable-state population with (a) the intensity and (b) the number of passes through the atomic beam of OP light changes.  The sub-linear increase of the metastable-state population with intensity indicates that the OP transition is highly saturated, i.e.\ $I_{op} \!\gg\! I_{sat,op} \!\simeq\! \SI{4}{\milli\watt/\square\centi\m}$.  The increase in metastable-state population with the number of passes of OP light through the chamber reflects the fact that, even at saturation, the number of photons scattered by by an atom within a single pass of the OP light is insufficient to guarantee optical pumping.}
    \label{fig:atom_number}
\end{figure}

Second, we find that the metastable atom flux at zero transverse velocity, i.e.\ the peak values in the transverse Doppler broadened distributions, is strongly enhanced in the optically pumped beam as compared to the unpumped beam.  The maximum enhancement factor we observe, as shown in Fig.\ \ref{fig:TDPs}a for a setup with four passes of maximum-intensity OP light through the chamber (see Fig.\ref{fig:atom_number}), is $\sim\! 40$. From geometry we ascertain that the ratio of atoms that pass through the OP light and the probe beam to those that pass only through the probe beam is 0.23. Taking into account this geometric factor, the enhancement we see among atoms that pass through the OP region, is $\sim\!170$. As discussed above, the maximum efficiency of optical pumping is estimated to be $\varepsilon^{\mathrm{op}}_\mathrm{max} \!=\! 0.5$ on the transition we have chosen.  By comparison, in the absence of optical pumping, the ratio of population in the metastable state to that in the ground state is measured to be $r_\mathrm{thermal} \!=\!1.1(2)\times 10^{-3}$. The maximum enhancement ratio expected for optical pumping is thus $\varepsilon^{\mathrm{op}}_\mathrm{max} / r_\mathrm{termal} \!=\! 450$.  The fact that we observe an enhancement ratio significantly smaller than this maximum may reflect an inefficiency in pumping atoms with fast longitudinal velocities owing to finite transit time and saturation (as discussed above), and also an over-estimate of $\varepsilon^{\mathrm{op}}_\mathrm{max}$ owing to uncertainty in the transition rates and branching ratios.

The efficiency of optical pumping is explored further in Fig.\ \ref{fig:atom_number}.  We find that the peak signal in the \SI{498}{\nm} wavelength fluorescence increases with the power of the OP light, indicating an increase in the density of zero-transverse-velocity atoms being pumped.  This increase is sub-linear, supporting the assumption that the OP transition is highly saturated. In this regime, we presume that as power is increased, a $\sqrt{s+1}$ increase in the velocity width of addressed atoms takes place as well as an effective increase in the area of our OP beam, thereby increasing the population of metastable atoms that reach the down-stream region probed by our fluorescence beam.  We observe also that the overall pumping efficiency increases when the OP beam is retro-reflected multiply through the chamber, and observation that supports the hypothesis that, for short interaction lengths, fast atoms are not pumped efficiently.

\section{Conclusion}
\label{sec:conclusion}

We now turn to the question of loading a magneto-optical trap (MOT) with a Ti-ball beam. As mentioned in the introduction, we find a metastable atomic flux density
of \SI{4.3(2)e9}{\per\s\per\square\centi\m} at a distance of \SI{2.55}{\cm} away from the center of the Ti-ball. The half-angle divergence of
the beam is
$\sim\!0.6\degree$ along the OP direction ($x$) and $\sim\!5\degree$ along the other transverse direction ($y$). We note that this is the flux density in the spatial region directly behind the small OP beams. Suppose we expand the OP beam waist along the direction transverse to the atom beam so that the entire output of the Ti-ball aperture is covered and the optical pumping provides no additional spatial filtering. Suppose also that additional power is supplied to maintain the same saturation as in our system.

Because of the narrow collimation along the OP direction, the flux density only falls off as $1/r$ with the distance away from the source (circumference law as opposed to surface area law). The flux density as a function of distance $r$ from the source is approximately $\Phi/A \approx\SI{1.10(5)e10}{\per\s\per\square\centi\m}\times\frac{1}{r}$.  Noting the measured forward velocity distribution of the optically pumped beam and considering a conservative MOT capture velocity of \SI{30}{\m/\s}, for a MOT with \SI{1}{\cm} beams located \SI{10}{\cm} downstream of the Ti-ball, we estimate a MOT loading rate of \SI{1.5e5}{\per\s}. For an optimistic capture velocity of \SI{100}{\m/\s}, the number is \SI{5.4e6}{\per\s}.

We note two crucial points about these loading rates. First, our measured forward velocity distribution and sublimation rate are consistent with a Ti-ball temperature in the range of 1345-\SI{1430}{\kelvin}. It will be straightforward to increase the temperature of the Ti-ball up to around \SI{1600}{\kelvin}, within specifications for operating the Ti-ball.  Achieving this in our setup would require a simple replacement of our copper electrical feedthroughs with a less thermally conductive feedthrough, e.g.\ using narrower-gauge steel rods.  At this higher temperature, with the Ti-Ball sublimating up to \SI{6}{\mg/\hour}, we expect all measures of beam flux and density to increase by a factor $\geq\!60$, with only slight variations in the beam's velocity distributions from the moderately higher source temperature. Additionally, at \SI{1345}{\kelvin} (\SI{1600}{\kelvin}), only $0.013\%$ ($0.011\%$) of the optically pumped beam has a forward velocity lower than \SI{30}{\m/\s} (here we are using the modified Maxwell-Boltzmann distribution which describes the forward velocity of the optically pumped beam). If a \SI{10}{\cm} Zeeman slower is added to the system to slow atoms at \SI{300}{\m/\s} to \SI{30}{\m/\s}, the population becomes $8\%$ ($7\%$). A Zeeman slower of \SI{1}{\m} gives $77\%$ ($70\%$).

In the case of a \SI{1600}{\kelvin} optically pumped Ti-ball feeding into a \SI{10}{\cm} Zeeman slower with a total distance of \SI{20}{\cm} from the beam source to a MOT region, we estimate a  MOT loading rate of \SI{2.3e9}{\per\s} assuming \SI{30}{\m/\s} capture velocity.

This work raises the possibility of using a Ti-ball-like source as the starting point for laser cooling other transition-metal elements.  Among the elements considered in Ref.\ \cite{eustice_laser_2020}, there are five other near-refractory elements with sublimation temperatures in the range produced by a Ti-ball-like device, these being Sc (with sublimation temperature of \SI{1064}{\kelvin}, at which the equilibrium vapor pressure reaches $10^{-8}$ mbar), Y (\SI{1236}{\kelvin}), V (\SI{1427}{\kelvin}), and Fe (\SI{1115}{\kelvin})\cite{alcock1984}.  A Ti-ball-like source, i.e.\ a solid mass of elemental material placed around an electrically heated tungsten filament, could be used for each of these to produce atomic beams.

\section*{Acknowledgements}

We thank Jesse Lopez in the Berkeley Student Machine Shop for his assistance in designing and building the Ti-ball holder, to Marianna Safronova and her group at the University of Delaware for assisting with the search for a proper optical pumping transition, and Ryan Rivers from the Berkeley Marvell Nanolab for his insight into developing a Ti beam source. This material is based upon work supported by the U.S. Department of Energy, Office of Science,
National Quantum Information Science Research Centers,
Quantum Systems Accelerator. Additional support is acknowledged from the ONR (Grants No. N00014-20-1-2513
and No. N00014-22-1-2280), and from the NSF (PHY-2012068).

\section*{Author Declarations}
\subsection*{Conflict of interest}
The authors have no conflicts to disclose.

\subsection*{Author contribution}
\noindent\textbf{Jackson Schrott}: Data curation (equal); Formal analysis (lead); Investigation (supporting); Methodology (lead); Software (lead); Validation (equal); Visualization (lead); Writing – original draft (equal); Writing – review \& editing (equal). \textbf{Diego Novoa}: Data curation (equal); Investigation (lead); Methodology (supporting); Resources (lead); Validation (equal); Visualization (supporting); Writing – original draft (equal). \textbf{Scott Eustice}: Conceptualization (equal); Formal analysis (supporting); Methodology (supporting); Project Administration (lead); Supervision (supporting); Validation (equal); Writing – review \& editing (equal). \textbf{Dan Stamper-Kurn}: Conceptualization (equal); Funding Acquisition (lead); Supervision (lead); Validation (equal); Writing – review \& editing (equal).

\section*{Data availability}
The data that support the findings of this study are available within the article and also from the corresponding authors upon reasonable request.

\bibliography{ti_ball_refs,allrefs_x2}% Produces the bibliography via BibTeX.

\appendix

\section{Modelling scattering rates}
\label{sec:appendix}

To infer a measurement of atomic density from collected fluorescence, a detailed understanding of the atomic scattering rate is required. A single atom prepared in the lower-energy state of a particular atomic transition, with an initial position in the beam aperture $\mathbf{r}_0$ and velocity $\mathbf{v}_0$, %\dmsk{Curious, why are you using the letter ``o'' as a subscript here?  Not sure what it signifies}\jack{Im using it to clarify that these are the velocity and position before the atom enters the probe beam and not the velocity/position that are functions of time as the atom passes through the beam and experiences the mechanical effects of the light. ie for an initial trajectory thru the beam defined by $\mathbf{v}_0$ and $\mathbf{r}_0$, the rate equations for $P^e(t)$ can be solved. When I integrate over $\mathbf{v}_0$ and $\mathbf{r}_0$, Im trying to make it clear that Im integrating over initial conditions to the initial value problem for $P^e(t)$, not integrating over the actual $v(t)$ and $r(t)$ that result from solving the time evolution. Can change the nomenclature if it would be helpful}
passing through a probe laser of Intensity $I(\mathbf{r})$, wavevector $\mathbf{k}$ and detuning $\Delta$ will fluoresce photons at a rate
\begin{align}
\Gamma = \gamma P^e(t)
\end{align}
%\dmsk{For future reference: I'm fixing several places where you've started a new paragraph after an equation, but where you shouldn't have}\jack{ty $\surd$}
where $\gamma=\sum_iA^{ki}$ is the total decay rate out of the excited state and $P^e(t)$ is the time-dependent population in the excited state induced by the probe field. The population, $P^e(t)$, is %\dmsk{Run-on sentence.  Rephrase}\jack{ $\surd$} 
the solution to a system of rate equations conditioned upon $\mathbf{r}_0$, $\mathbf{v}_0$, $I(\mathbf{r})$, $\mathbf{k}$, and $\Delta$. If the lower state of the transition is denoted $g$, the excited state $e$, other dark states $d$, and supposing the states are fine structure levels with magnetic sublevels $m_J$, the populations obey the following rate equations
\begin{align}
\nonumber \dot{P}^e_{m'_J} &= \sum_{m_J} R^{ge}_{m_J,m'_J}(P^g_{m_J} -P^e_{m'_J}) - (A^{e g} + A^{e d})P^e_{m'_J} \\
\nonumber \dot{P}^g_{m_J} &= -\sum_{m'_J} R^{ge}_{m_J,m'_J}(P^g_{m_J} -P^e_{m'_J}) + \sum_{m'_J}A^{e g}_{m_J,m'_J}P^e_{m'_J} \\
\dot{P}^d_{m_J} &= \sum_{m'_J}A^{ed}_{m_J,m'_J}P^e_{m'_J}
\label{eqn:rate_eqns}
\end{align}

Here, $A^{ki}$ are Einstein coefficients between levels, and $R^{eg}_{m_J,m'_J}$ is the rate of optical pumping between magnetic sublevels of the lower and upper state and is given by
\begin{align}
R^{ge}_{m_J,m'_J} = \frac{16d_{eg}^2|\langle J_g m_J 1 q|J_e m'_J \rangle|^2I(\mathbf{r})}{c\epsilon_0 \hbar^2 A^{eg}\Big(1+4\Big(\frac{\Delta_{m_J,m'_J}-\mathbf{k} \cdot \mathbf{v}}{A^{eg}}\Big)^2\Big)}
\label{eqn:Ropt}
\end{align}
where $d_{eg}$ is the reduced dipole matrix element between the $e$ and $g$ sates, and $\Delta_{m_J,m'_J}$ are the Zeeman shifted detunings.  The index $q$ indicates %\dmsk{I'm fixing a few places where you're starting sentences with math symbols.  No no.}\jack{ty $\surd$}
the polarization of light in the spherical basis. The rate equations capture the effects of power broadening, Doppler broadening, optical pumping among $m_J$ states, and scattering to dark states on the total photo emission rate. It is implicitly assumed that $\mathbf{r}$ and $\mathbf{v}$ are themselves functions of time with the component $\mathbf{v}$ along $\mathbf{k}$ growing in proportion to the integral of the instantaneous scattering rate. Therefore the mechanical effect of the probe light on the atom, which accelerates the atom so its Doppler shift is pushed out of resonance with the probe beam, is also captured.

Consider a beam of atoms, interacting with a fluorescence probe as shown in Figure \ref{fig:apparatus}. We define a three-dimensional region of interest (VROI) formed by the boundaries of the camera region of interest (ROI) in the object plane and the cross-sectional area of the probe beam. We consider the situation where the beam radius and camera ROI are small compared to the extent of the atom beam so the steady-state density of atoms in the VROI is approximately uniform.

The average rate at which an atom with initial position $\mathbf{r}_0$ and velocity $\mathbf{v}_0$ fluoresces photons during during its time in the VROI is $\Gamma_{\mathrm{avg}}=\frac{\gamma}{\tau}\int_0^{\tau}P^e(t, \mathbf{r}_0, \mathbf{v}_0, I(\mathbf{r}), \mathbf{k}, \Delta)dt$, where $\tau$ is the time between when the atom enters and exits the VRIO and is itself a function of $\mathbf{r}_0$ and velocity $\mathbf{v}_0$.

If the atoms that pass through the VROI have initial position and velocity distributions $d(\mathbf{r}_0)$ and $f(\mathbf{v}_0)$, then the rate at which atoms are detected by the camera is
\begin{align}
\nonumber R_{\mathrm{det}}=\epsilon\eta Vn_a\iint \Big\{ \frac{\gamma}{\tau}\int_0^{\tau}P^e&(t, \mathbf{r}_0, \mathbf{v}_0, I(\mathbf{r}), \mathbf{k}, \Delta)dt\Big\} \\
&\times d(\mathbf{r}_0)f(\mathbf{v}_0)d^2r_0d^3v_0
\end{align}
Here, $V$ is the volume of the VROI, $n_a$ is the steady state density of atoms in the state under consideration, $\epsilon$ quantifies the total photon collection efficiency of the imaging system, including numerical aperture, optical losses, and detector quantum efficiency, and $\eta$ quantifies the fraction of fluoresced photons that are emitted within the optical bandwidth of our detector, i.e., in our setup, the branching ratio for decay from the atomic excited state that cycles atoms to their original lower-energy state. We note again that $\tau$ is determined by $\mathbf{r}_0$ and $\hat{\mathbf{v}}_0$ and must be included under the integral.

From Equations \ref{eqn:rate_eqns} and \ref{eqn:Ropt}, it can be seen that $P^e(t)$ depends distinctly on the components of $\mathbf{v}_0$ that are parallel and perpendicular to $\mathbf{k}$. Denoting these $v_{\parallel}$ and $v_{\perp}$, we see $v_{\parallel}$ dictates the Doppler shift of the atom and is changed by the interaction with the probe light,  whereas $v_{\perp}$ may only affect the interaction time of the atom with the beam and is not altered by the probe (ignoring small momentum kicks from isotropic scattering). Consider a configuration where atoms always enter and exit the VROI through the sides of a Gaussian probe beam as opposed to the planes defined by the boundary of the camera ROI in the object plane. Then the interaction time of the atom in the VROI is $\tau=l_c/v_{\perp}$, where $l_c$ is the chord the atom's path makes with the circular beam profile. For instance, this would be the case for a probe beam directed transversely to the atomic beam. In this configuration, $l_c$ uniquely determines the intensity profile the atom sees. Considering this situation and assuming a fixed radial profile $I(\mathbf{r})$ and wavevector $\mathbf{k}$, we can change variables and write
\begin{align}
\nonumber R_{\mathrm{det}}(\Delta) = \epsilon\eta Vn_{a}\gamma&\iiiint\Big\{\frac{v_{\perp}}{l_c}P^e(t, v_{\parallel}, v_{\perp}, l_{c},\Delta)\times\\ 
&f(v_{\parallel})g(v_{\perp})h(l_{c})\Big\}dt\,dv_{\parallel}\,dv_{\perp}\,dl_{c}
\label{eqn:Ndet_mod}
\end{align}

It can be seen from solving the rate equations that, to a very good approximation, $P^e$ is peaked at $v_{\parallel}=\Delta/k$. If the power-broadened line width of the atomic transition is much narrower than the Doppler width of the beam, $f(v_{\parallel})$ may be considered slowly varying in $v_{\parallel}$ compared to $P^e$, and be evaluated at $v_{\parallel}=\Delta/k$ and pulled out of the integral. Additionally, from Equation \ref{eqn:Ropt}, it can be seen that $P^e$ depends only on the difference $\Delta - v_{\parallel}k$, and not on $\Delta$ and $v_{\parallel}$ separately. Therefore making the coordinate change $v'_{\parallel} = v_{\parallel} - \Delta/k$ and then dropping the prime, the dependence on $\Delta$ can be dropped in the integral. The velocity distribution parallel to the beam can then be related to the number of photons detected at a particular detuning by
\begin{align}
\nonumber &n_{a}f(\Delta/k) \approx \frac{R_{\mathrm{det}} (\Delta)}{\epsilon\eta V\gamma}\xi^{-1} \\
\nonumber &\xi = \iiiint\Big\{\frac{v_{\perp}}{l_c}P^e(t, v_{\parallel}, v_{\perp}, l_{c})\times\\ 
&\quad\quad\quad\quad\quad\quad\quad
g(v_{\perp})h(l_{c})\Big\}dt\,dv_{\parallel}\,dv_{\perp}\,dl_{c}
\label{eqn:Nfparallel}
\end{align}

The rate $R_{\mathrm{det}}(\Delta)$ is measured in the lab by tuning the probe laser over a range of $\Delta$ and measuring the number of photons detected within the camera ROI during an exposure time $t_{\mathrm{ex}}$. From this, $f(v_{\parallel})$ can be found by normalizing the resulting distribution. To extract $n_a$ in addition to $f(v_{\parallel})$, $\xi$ must be calculated. $\xi$ can be evaluated first by numerically solving for $P^e$, and integrating with $g(v_{\perp})$ and $h(l_c)$. Moreover, $g(v_{\perp})$ can be measured and $h(l_c)$ can be determined from geometry. We reiterate that Eq.\ \ref{eqn:Nfparallel} applies to the situation where atoms enter and exit the VROI out of the sides of the beam, i.e. for a transverse probe. A similar expression can be derived for the situation where atoms enter and exit the VROI through the planes defined by the boundaries of the camera ROI in the object plane instead.

\end{document}